\documentclass[aip,
rsi,%
 amsmath,amssymb,
 reprint,%
]{revtex4-1}

\usepackage{graphicx}
\usepackage{dcolumn}
\usepackage{bm}
\usepackage[version=3]{mhchem}
\usepackage{threeparttable}
\usepackage{titlesec}
\usepackage{xcolor}
\usepackage{ulem}
\usepackage{xr}
\usepackage{natbib}

\begin{document}

\title{The MD17 Datasets from the Perspective of Datasets for Gas-Phase ``Small" Molecule Potentials}
\date{\today}
\author{Joel M. Bowman}
\email{jmbowma@emory.edu}
\affiliation{Department of Chemistry and Cherry L. Emerson Center for Scientific Computation, Emory University, Atlanta, Georgia 30322, U.S.A.}
\author{Chen Qu}
\affiliation{Independent Researcher, Toronto, Canada}
\author{Riccardo Conte}
\email{riccardo.conte1@unimi.it}
\affiliation{Dipartimento di Chimica, Universit\`{a} Degli Studi di Milano, via Golgi 19, 20133 Milano, Italy}
\author{Apurba Nandi}
\affiliation{Department of Chemistry and Cherry L. Emerson Center for Scientific Computation, Emory University, Atlanta, Georgia 30322, U.S.A.}
\author{Paul L. Houston}
\email{plh2@cornell.edu}
\affiliation{Department of Chemistry and Chemical Biology, Cornell University, Ithaca, New York
14853, U.S.A. and Department of Chemistry and Biochemistry, Georgia Institute of
Technology, Atlanta, Georgia 30332, U.S.A}
\author{Qi Yu}
\affiliation{Department of Chemistry Yale University, New Haven, Connecticut 06520, U.S.A.}


\begin{abstract}
There has been great progress in developing methods for machine-learned potential energy surfaces.  There have also been important assessments of these methods by comparing so-called learning curves on datasets of electronic energies and forces, notably the MD17 database.  The dataset for each molecule in this database generally consists of tens of thousands of energies and forces obtained from DFT direct dynamics at 500 K.  We contrast the datasets from this database for three ``small" molecules, ethanol, malonaldehyde, and glycine, with datasets we have generated with specific targets for the PESs in mind: a rigorous calculation of the zero-point energy and wavefunction,  the tunneling splitting in malonaldehyde and in the case of glycine a description of all eight low-lying conformers. We found that the MD17 datasets are too limited for these targets.  We also examine recent datasets for several PESs that describe small-molecule but complex chemical reactions. Finally, we introduce a new database,``QM-22", which contains datasets of  molecules ranging from 4 to 15 atoms that extend to  high energies and a large span of configurations.
\end{abstract}
\maketitle

\section{Introduction}
There has been dramatic progress in Machine-Learned Potentials (MLPs) over the past 10--15 years with many Reviews and Perspectives, and five Perspectives in J. Chem. Phys.\cite{persp0, persp1, persp2, persp3, persp4} The most recent one\cite{persp4} is a survey of numerous approaches for MLPs for materials research. To quote from that paper, ``Our discussion will be primarily about the use of interatomic potential models to model materials, ... We will focus on models that construct a continuous potential energy surface that allows for the calculation of forces and molecular dynamics (MD) simulations." As stated, the focus is on MD simulations, and indeed that has been the major focus of the work in materials. Also note a very recent Perspective on methods and datasets for many molecules and materials with a focus on structures.\cite{rupp22}

There has also been major work developing MLPs for isolated molecules in the gas-phase, the vast majority of which contain hydrogens atoms and thus require some treatment of nuclear quantum effects. Also, the focus is on molecular properties, e.g., isomerization, conformers, reaction dynamics, spectroscopy, etc. This contrasts to the focus of materials research, much of which deals with heavy atoms and is aimed at bulk thermal properties and others for which classical MD simulations are suitable.

The focus of this Perspective is to examine the datasets typically seen in these two areas.  For this purpose we consider several from the MD17 datasets\cite{MD17,Lilienfeld19} and ones from our work for the same molecules.  

The molecules at the center of much of the gas-phase work are sometime referred to as ``small" and so we adopt that terminology here and in the title.  However, as we will show the datasets for this class of molecules can be large.  Also we will point out that this class of molecules now routinely includes ones with 15 atoms. And in this regard the spheres of materials and isolated molecules are beginning to overlap and so there is an opportunity to look at datasets from these two spheres. We do that here.
 
A secondary but important common focus of both areas worth mentioning is the assessment of MLP methods in terms of precision, computational efficiency, etc. Assessments of these various methods have begun to appear.\cite{paescomps,PIP-GP,Tkatchjcp,dralchemsci,PIPSJCP22} The paper by Dral and co-workers\cite{dralchemsci} is particularly noteworthy as it examines the performance of ML methods GAP-SOAP,\cite{GP-2015-1} ANI,\cite{AN1} DPMD,\cite{dpmd2018} sGDML,\cite{Tkatchjcp} PhysNet,\cite{PhysNet} KREG,\cite{KREG} pKREG,\cite{pKREG} and KRR-CM\cite{KREG} for fits for ethanol and other molecules, all using the extended MD17 dataset of energies and forces\cite{Christensen_2020}. This popular dataset and the predecessor MD17\cite{MD17} provide energies and forces for 10 molecules obtained from direct-(classical)dynamics calculations at 500 K.  We recently added and assessed the PIP method for ethanol using the MD17 dataset, and\cite{PIPSJCP22} that is how we came to use that dataset. 

Specifically we examine datasets for ethanol, malonaldeyde and glycine. These are 9 and 10-atom molecules, so glycine could be considered borderline ``large", but that's not of importance.  We note that the glycine dataset that we label as ``MD17" is not in the MD17 database.  It was taken from ref.\citenum{tkatgly21}, where the protocol for MD17 datasets was used  with an enhancement of considering two conformers and a path between them.  More details are given below.


Before proceeding, and at the risk of stating the obvious, it is worth clarifying what is meant here by a potential energy surface (PES).  By this we mean a full-dimensional, faithful and precise representation of the adiabatic electronic energy (including nuclear repulsion) as a function of nuclear configuration.

Our group has been active for some time developing MLPs using Permutationally Invariant Polynomials (PIPs)\cite{Braams2009, msachen, ARPC2018, robo20, czakofa_21, paespips_21, jasperpips_21, greedypip_21, PIPSJCP22} for ``small molecules"; some early examples include \ce{CH5+} and \ce{H5O2+}, malonaldehyde and the 10-atom formic acid dimer.\cite{fad}  More recently, we have reported PIP PESs for 10-15 atom molecules, listed in Table \ref{tab:DMC} using enhancements to the PIP basis.\cite{chen2019, conte20, HoustonConteQuBowman2020} The PIPs method was recently evaluated\cite{PIPSJCP22} against the ML methods assessed in ref.\citenum{dralchemsci}. (PIPs were shown to be as precise as the most precise methods but to run much faster). 

The general target of our PESs is to be accurate at large enough energies to enable both rigorous diffusion Monte Carlo (DMC) calculations  of the zero-point energy and wavefunction\cite{mccoy04, mccoy2005full, McCoy2006, wanghex} as well as general quantum vibrational and scattering calculations and also  quasiclassical trajectory (QCT) calculations of chemical reaction dynamics.\cite{bowman11,cbfeature}

Focusing  on DMC calculations, we give in Table \ref{tab:DMC} DMC zero-point energies of a subset of molecules for which PIP PESs have been reported. The energies span a large range (from around 28 to 72 kcal/mol) depending on the size of the molecule, although the variation is not monotonic with the number of atoms.  Note also that all of these molecules have H atoms as the most common atom and this accounts in a major way for the large ZPEs.  The DMC calculation also provides the zero-point wavefunction from which all observable properties of this 0 K state can be derived. Before proceeding it is perhaps worth commenting on these rigorous 0 K energies and wavefunction and those from a normal-mode analysis.  Of course a full-dimensional PES is not needed for this analysis, which  actually provides a good approximation to these DMC energies.  However, the separable harmonic-oscillator model of the molecular motion has many well-known limitations, which can only be surmounted with a realistic PES that includes anharmonicity and mode-coupling.  And of course this is a motivation for MLPs.

In the next section we discuss a recent PIP PES for \ce{CH4}\cite{NandiQuBowman2019} as a primer for the investigations that follow for some of the larger molecules included in Table \ref{tab:DMC}. 

\begin{table}[htbp!]
\caption{Quantum zero-point energies (ZPEs) in kcal/mol and cm$^{-1}$) of indicated molecules from diffusion Monte Carlo calculation on PIP PESs. In \textbf{bold} the systems discussed in the text.}
\label{tab:DMC}

\begin{threeparttable}
	\begin{tabular*}{\columnwidth}{@{\extracolsep{\fill}}lcc}
	\hline
	\hline\noalign{\smallskip}
     Molecule &  ZPE (kcal/mol) &  ZPE (cm$^{-1}$)   \\
	\noalign{\smallskip}\hline\noalign{\smallskip}
    \textbf{Methane} \ce{CH4}  &27.82 & {  }9 730  \\
    Methonium \ce{CH5+}  &   31.38 &10 975   \\
    Water dimer \ce{(H2O)2}  &28.30&{  }9 898   \\
     Zundel Cation \ce{H5O2+}  &35.43& 12 393  \\
     \textbf{Ethanol} \ce{CH3CH2OH}  &49.58& 17 339 \\
     \textbf{Malonaldehyde} \ce{C3H4O2} &41.97&  14 678 \\
     \textbf{Glycine} \ce{C2H5NO2}  &49.04& 17 151  \\
     $N$-methyl acetamide \ce{C3H7NO} &62.63  & 21 905  \\
     Tropolone \ce{C7H6O2}  &71.77& 25 100  \\
   
	\noalign{\smallskip}\hline
	\hline
	\end{tabular*}
\end{threeparttable}
\end{table}

\section{A PES primer: Methane}
\begin{figure}[htbp!]
    \centering
    \includegraphics[width=1.0\columnwidth]{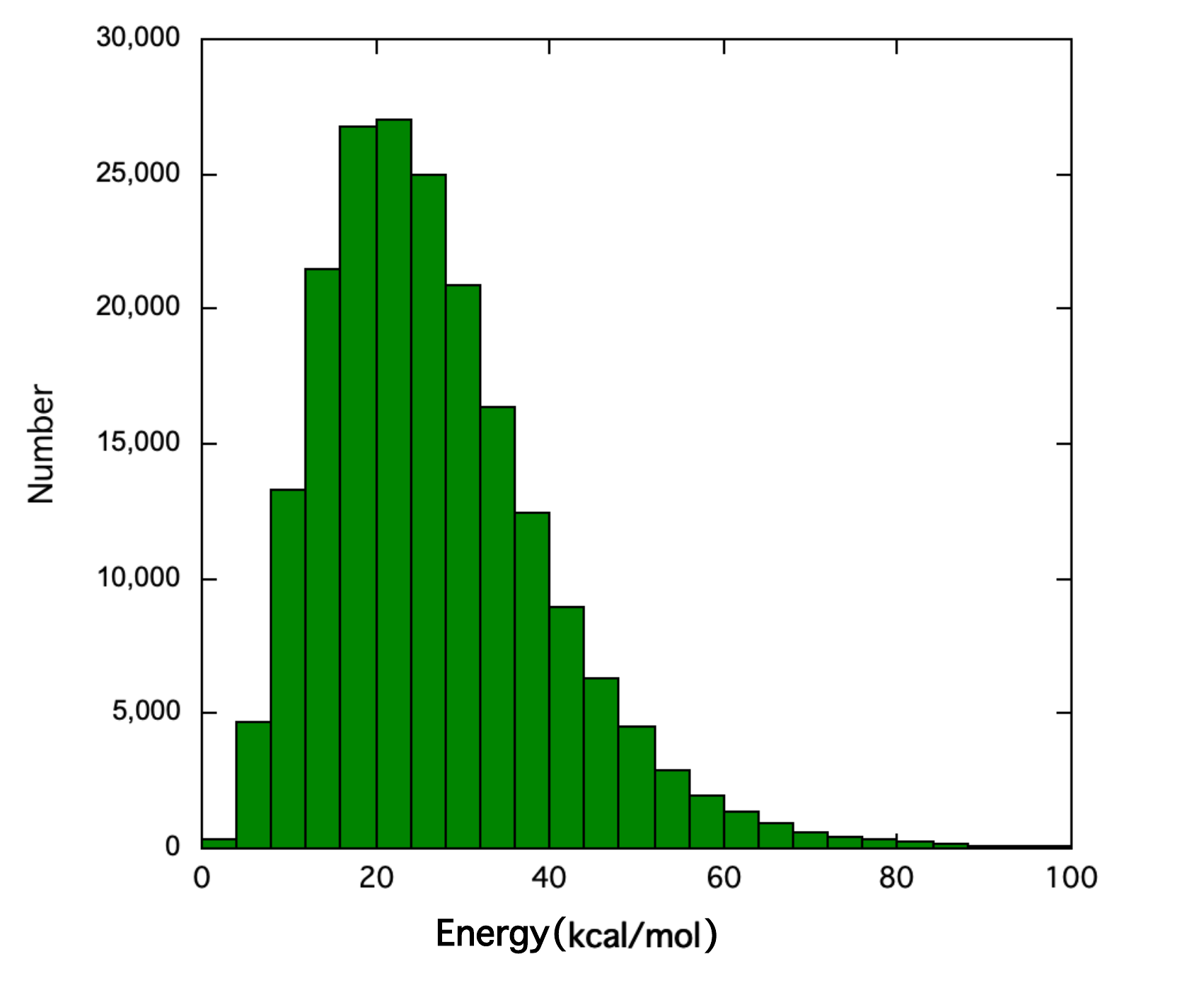}
    \caption{Histogram of methane potential energies sampled by the DMC zero-point wavefunction}
    \label{fig:histofenergies}
\end{figure}

An instructive place to begin is with a rigorous DMC calculation of zero-point energy of \ce{CH4}.  We have done this using a demonstration PES and B3LYP density functional theory energies and gradients for the fit.\cite{NandiQuBowman2019}  As usual, we probed for ``holes'', i.e., regions of unphysically low energy, using DMC calculations. These were found in initial fits and then ultimately eliminated by adding data in the dataset to be fitted.  The ZPE obtained with this method is 27.8 kcal/mol (9730 cm$^{-1}$) with a statistical uncertainty of several wavenumbers. (We recently reported DMC as general method, possibly with fictitious masses, to explore a PES for holes.\cite{li2021}) 
An interesting question to ask is: What is the distribution of potential energies sampled by the corresponding DMC wavefunction?  The answer is shown in Fig.\ref{fig:histofenergies}.  What we see is that this distribution is broad and peaks at a value near the ZPE, but extends significantly beyond that value.  This is exactly the qualitative behavior we expect to see for a wavefunction dominated by H atoms \textemdash ~that is, extension of the wavefunction significantly into the classically forbidden region.

So, clearly the PES used for this calculation needs to extend (faithfully) to energies higher than the ZPE.  In fact the PES used was a fit to data up to roughly 43 kcal/mol (15~000 cm$^{-1}$) which is about 1.5 times the ZPE.  The potential values above 43 kcal/mol shown in Fig.\ref{fig:histofenergies} are thus extrapolations of the PES to energies higher than the data used in the fit.   
With this example in mind, which has a focus on the rigorous ground vibrational state wavefunction, we examine a number of larger examples and datasets for PES fits which are taken from the literature.

\section{Case studies of three data sets from MD17, and our work}
Now we present three case studies that illustrate the points we are making. In all cases the dataset from our group is denoted by acronyms of the authors' last names. The data from MD17, obtained from DFT direct-dynamics run at 500 K, are labeled using that term.\cite{MD17,Lilienfeld19} This approach, i.e., using DFT direct-dynamics at thermal energies, perhaps as high as 1000 K, is commonly done in the field to generate data for MLPs of a given molecule.  We also use direct-dynamics as one means of generating configurations; however, at a number of total energies, including high energies.

\subsection*{Malonaldehyde}
The tunneling splitting of the proton transfer in malonaldehyde is perhaps the most studied example of proton transfer governed by a barrier.  The literature is extensive but of not direct relevance here. However, the interested reader is directed to some recent articles.\cite{malon, malonexph, tewmalon} A PIP ML PES to describe the proton transfer quantitatively (barrier height of 4.1 kcal/mol) used an extensive dataset, which exceeded the zero-point energy of roughly 42 kcal/mol by a factor of about 1.5, and was reported in 2008.\cite{malon} A newer ML (LASSO) PES was reported in ref. \citenum{tewmalon} with a slightly different barrier height. The dataset for that PES extended to roughly 72 kcal/mol, which again is significantly above the ZPE.

\begin{figure}[htbp!]
    \centering
    \includegraphics[width=1.0\columnwidth]{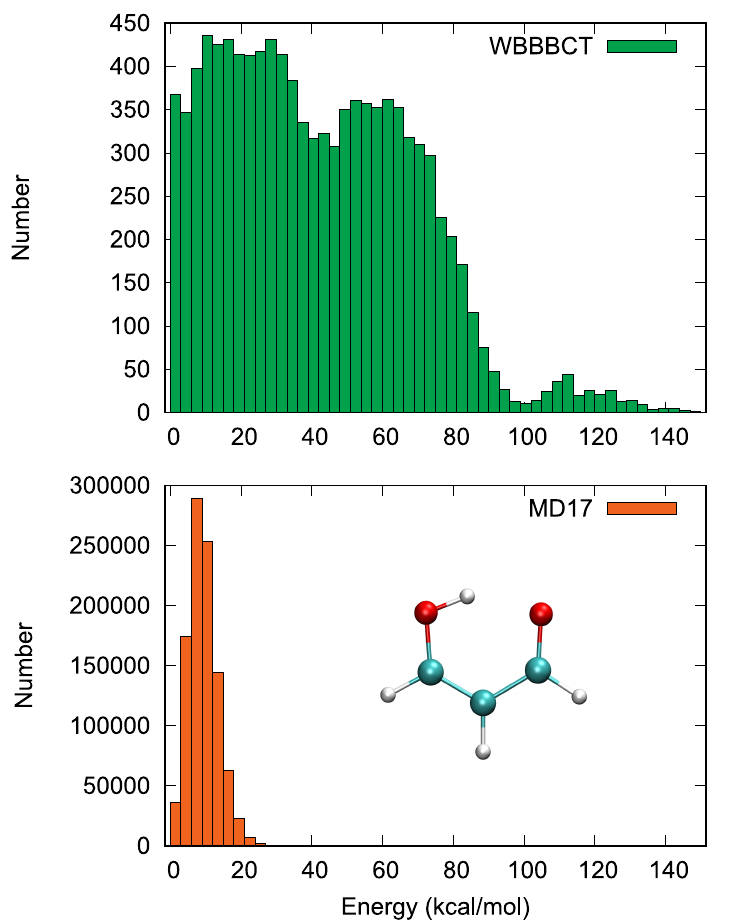}
    \caption{Histogram of malonaldehyde electronic energies from ref. \onlinecite{malon} and MD17\cite{MD17} datasets. The WBBBCT dataset is reproduced from Wang et al. J. Chem. Phys. \textbf{128}, 224314 (2008) with the permission of AIP Publishing.}
    \label{fig:malon}
\end{figure} 

The datasets of energies from MD17 dataset and Wang et al.\cite{malon} are given in Fig. \ref{fig:malon}. As seen, the former one is limited to a maximum energy of around 30 kcal/mol whereas the latter extends to much higher energies.  The structure in this dataset comes from several samples, details of which are given elsewhere.\cite{malon}  This is an important aspect of generating datasets for MLPs and we defer comments on this to the next section. As seen, the range for most of the energies extends to roughly 100 kcal/mol, which is 2.4 times the ZPE. There is a smaller set of energies that extend to 140 kcal/mol.  These high energies beyond the ZPE were added so that quantum calculations based on the Vibrational Self-Consistent Field and Configuration Interaction (VSCF/VCI)\cite{malon} and  Multi-Configuration Time-Dependent Hartree (MCTDH)\cite{manthemalon, meyermalon} methods could be done without encountering holes.  The methods were used to finally obtain the ``right answers", i.e., accurate tunneling splittings for malonaldeyde and d$_1$-malonaldehyde ``for the right reasons".

\subsection*{Ethanol}
Ethanol is an important case to consider next for several reasons.  First, it was the focus of two studies assessing numerous MLP methods, mentioned already.\cite{dralchemsci, PIPSJCP22}  These studies were based on the MD17 dataset.  Secondly, it is scientifically of fundamental interest as it has two nearly isoenergetic conformers (trans and gauche) and two different methyl rotors.  It is also of great applied interest in fields as diverse as combustion and astrochemistry.  

\begin{figure}[htbp!]
    \centering
    \includegraphics[width=1.0\columnwidth]{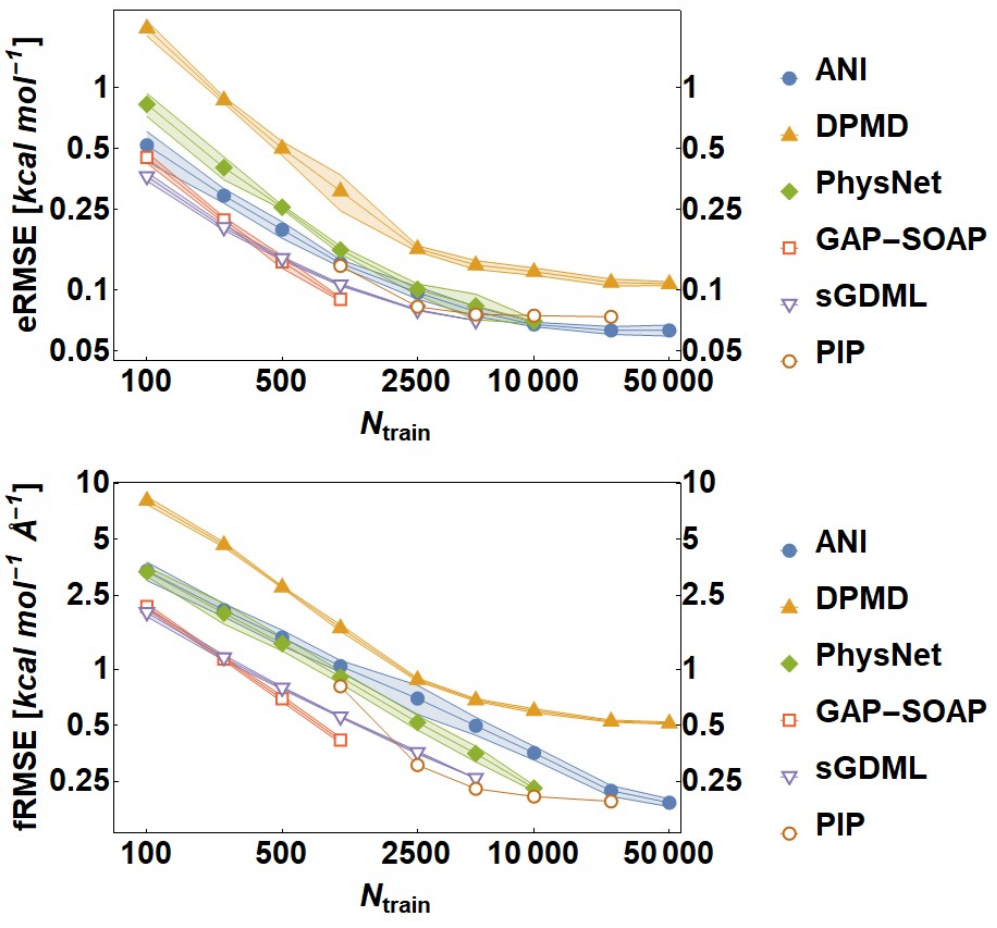}
    \caption{Ethanol PES learning curves. Reproduced from Houston et al. J. Chem. Phys. \textbf{156}, 044120 (2022), with the permission of AIP Publishing.}
    \label{fig:ethanol}
\end{figure}

In the recent paper from our group, \cite{PIPSJCP22} we examined learning curves for several PIP fits to the MD17 data set following the apporach of an earlier study,\cite{dralchemsci} where learning curves for all methods except the PIP one were reported.  One selected plot of these is given Fig. \ref{fig:ethanol}.  As seen, the performance of most methods is very good with the PIP one showing excellent performance.  In addition the computational speed of the PIP fits was shown to be factors of around 10 to 100 faster than other methods.  Details are given in ref. \citenum{PIPSJCP22}.  

\begin{figure}[htbp!]
    \includegraphics[width=\columnwidth]{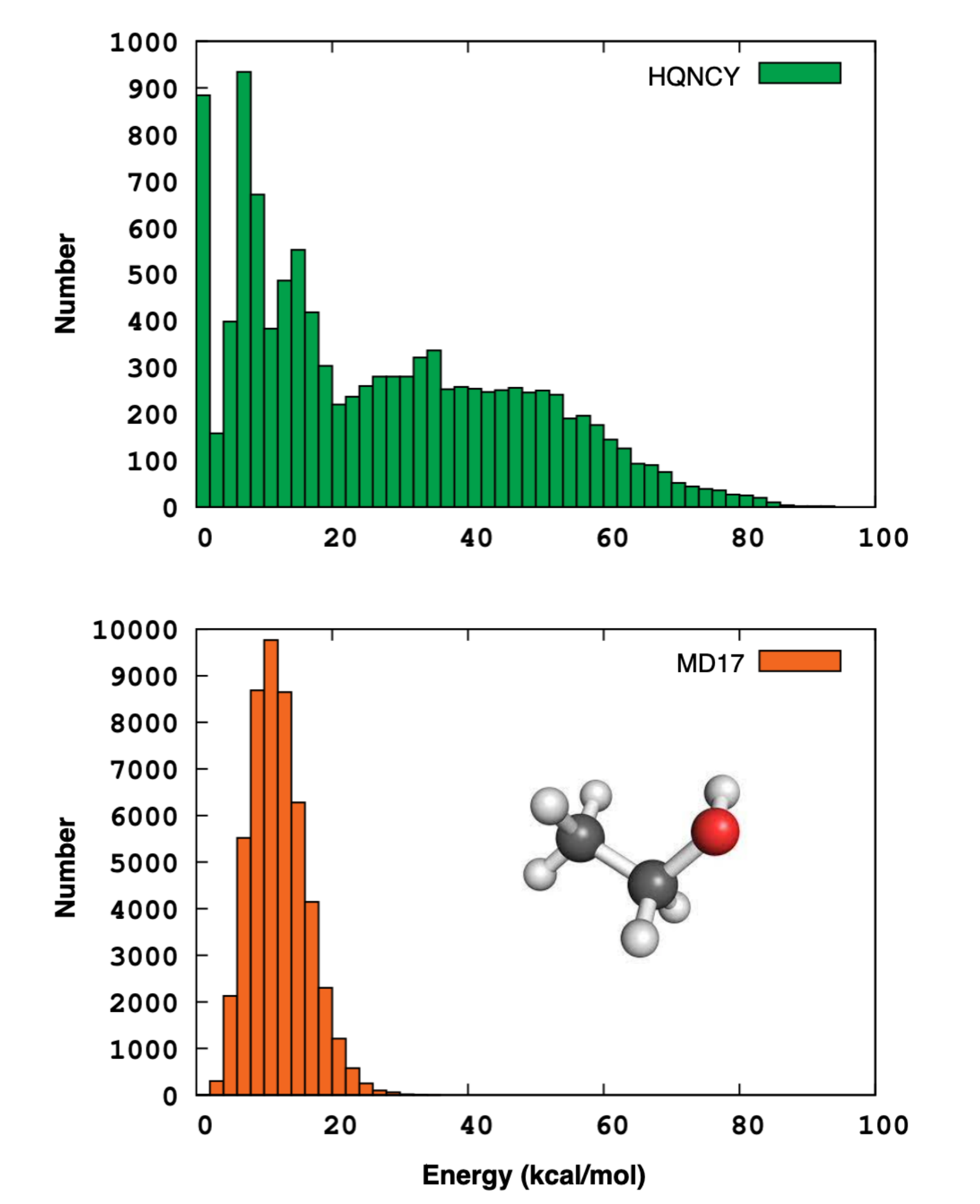}
    \caption{Histograms of ethanol electronic energies (kcal/mol)  from ref.\citenum{PIPSJCP22} and MD17\cite{MD17} datasets.}
        \label{fig:dataset_comparison}
\end{figure}

The MD17 energy distribution for ethanol is shown in Fig. \ref{fig:dataset_comparison} along with the dataset we recently reported.\cite{PIPSJCP22} We were motivated to extend the MD17 data set after we found many holes in the precise PIP fit to that dataset in DMC calculations. This new dataset was generated at the B3LYP/6-311+G(d,p) level of theory using our usual protocol, i.e., direct-dynamics at a number of total energies and having much larger coverage of configuration space and energies than the MD17 set. These trajectories were propagated for 20 000 time steps of 5.0 a.u. (about 0.12 fs) and with total energies of 1000, 5000, 10~000, 20~000, 30~000, and 40~000 cm$^{-1}$. A total of 11 trajectories were calculated; one trajectory at the total energy of 1000 cm$^{-1}$ and two trajectories for each remaining total energies. The final data set consists of 11~000 energies and corresponding 297~000 forces. 
The extended dataset was fit with the same large PIP basis used for the assessment of PESs based on the MD17 dataset. This new PES was used successfully in DMC and semi-classical (SC) calculations of the ZPE. Thus, this new PES is mostly an example of the ease with which PESs for a molecule with 9 atoms and two methyl rotors can be developed.

\subsection*{Glycine}
\begin{figure}[htbp!]
    \centering
    \includegraphics[width=1.\columnwidth]{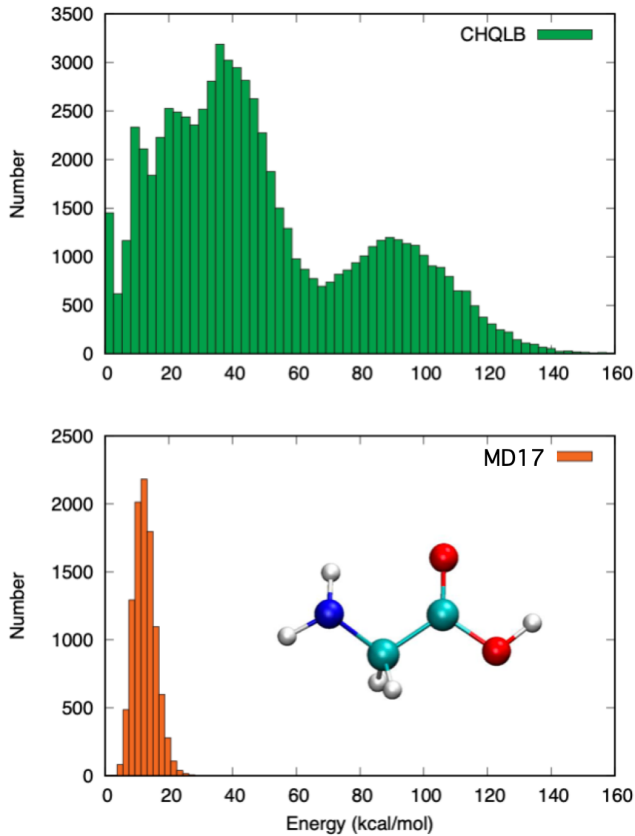}
    \caption{Histogram of glycine electronic energies (kcal/mol)  from ref.\citenum{glycine2019} and MD17\cite{MD17} datasets.}
    \label{fig:histogly}
\end{figure}

Next we consider 10-atom glycine. Histograms of the MD17 energies and the one we recently reported\cite{glycine2019} are shown in Fig. \ref{fig:histogly}. We do note that in this case the ``MD17'' dataset was expanded to include conformer 3 and a reaction path between the lowest energy conformer 1 and conformer 3. No other conformers were included in the database.

Once again there is major difference between the two datasets. Our database for the glycine PES was constructed using DFT with B3LYP hybrid density functional and an aug-cc-pVDZ basis set.\cite{glycine2019} Both the energies and gradients were calculated and then fitted using the PIP method. The final database included 70~099 geometries and the fit was performed on all energies, and on those gradients associated with the 20~000 geometries with lower energies. The fit was also inverse-energy weighted and gradients were given 1/3 of the weights of the energies.

\begin{figure}[htbp!]
    \centering
    \includegraphics[width=1.0\columnwidth]{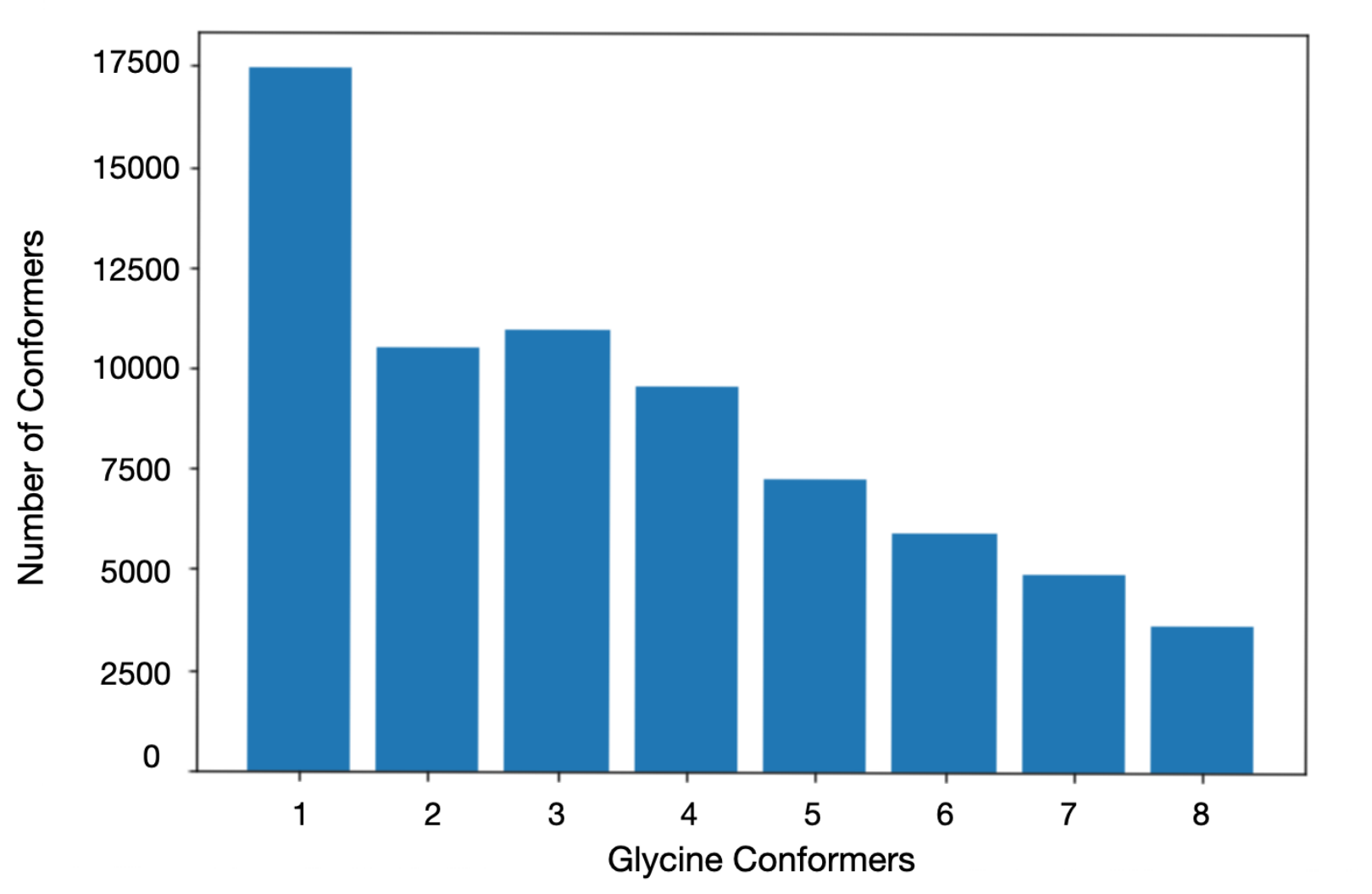}
    \caption{Histogram of distribution of conformers of glycine for the dataset based on assigning a given configuration to the closest conformer as explained in ref. \citenum{glycine2019}.}
    \label{fig:glyclustering}
\end{figure}

Geometries for the database were chosen by an iterative process by first undertaking classical sampling using direct-dynamics, then performing a preliminary fit, and finally by adding points from randomly generated grids centered on the stationary points of the preliminary surface. To assess the reliability of the surface produced by this classical sampling, we used quantum DMC, which frequently revealed holes in the surface. These were ``filled" by adding configurations at the hole configurations.  Ultimately, the dataset consisted of 70~000 configurations, including 8 conformers and 15 saddle points between them. Fig. \ref{fig:glyclustering} shows how geometries in the database are distributed according to their nearest conformer. From the Figure it is evident that the global minimum (Conformer 1) has been sampled in more detail but the coverage is pretty uniform for all conformers including higher-energy ones. Table \ref{tab: confs} details the energetics of the 8 conformers.


\begin{table}[htp!]
\centering
\caption{Energies relative to the global minimum (in kcal/mol unless otherwise indicated in the Table) of each conformer from the PES or indicated electronic structure theories. All data is from ref.\citenum{glycine2019} except the CCSD(T)-F12 numbers which are from ref.\citenum{Orjan2020}}
\label{tab: confs}

\begin{threeparttable}
    \begin{tabular*}{\columnwidth}{@{\extracolsep{\fill}}lcccc}
    \hline
    \hline\noalign{\smallskip}
    Conformer & PES (cm$^{-1}$) & PES & B3LYP & CCSD(T)-F12 \\
    \noalign{\smallskip}\hline\noalign{\smallskip}
    Conf 1 &    0.0 & 0.00 & 0.00 & 0.00 \\
    Conf 2 &  205.1 & 0.59 & 0.58 & 0.68 \\
    Conf 3 &  577.1 & 1.65 & 1.64 & 1.73 \\
    Conf 4 &  450.4 & 1.29 & 1.27 & 1.23 \\
    Conf 5 &  945.9 & 2.70 & 2.61 & 2.62 \\
    Conf 6 & 1719.4 & 4.92 & 4.91 & 4.80 \\
    Conf 7 & 2043.6 & 5.84 & 5.84 & 5.89 \\
    Conf 8 & 2174.3 & 6.22 & 6.25 & 6.06 \\
    \noalign{\smallskip}\hline
    \hline
    \end{tabular*}

\end{threeparttable}
\end{table}

\begin{figure}[htbp!]
    \centering
    \includegraphics[width=1.0\columnwidth]{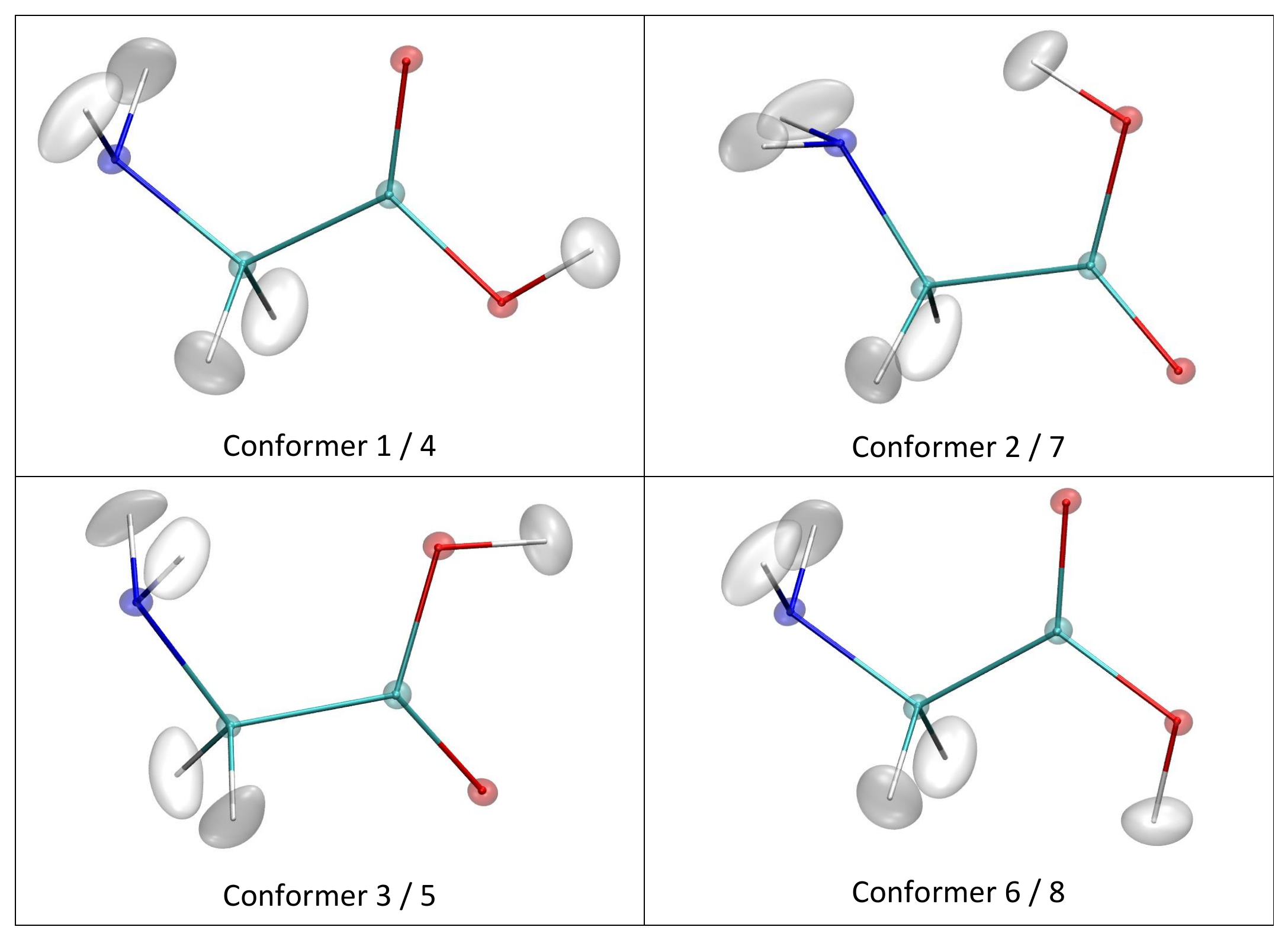}
    \caption{Isosurface plots of diffusion Monte Carlo ground-state vibrational wavefunctions for glycine. Reproduced from Conte et al., J. Chem. Phys. \textbf{153}, 244301 (2020), with the permission of AlP Publishing.}
    \label{fig:glycineconf}
\end{figure}

DMC calculations were performed by initiating walkers at the minimum of each conformer. The zero-point energies and wavefunctions were determined.\cite{conte_glycine20}  Isosurface plots of the wavefunctions are shown in Figure \ref{fig:glycineconf} where it is clear that the 8 conformers can be grouped into pairs described by energetically isolated asymmetric double wells. The corresponding ZPEs were obtained and indeed  these group into 4 values instead of 8. This important finding, i.e., that there are in fact 4 distinct conformers and not 8, required the use of unrestricted DMC calculations.
 
Semiclassical calculations were performed on the PES by means of the adiabatically switched semiclassical initial value representation method (AS SCIVR)\cite{Conte_Ceotto_ASSCIVR_2019, Botti_Conte_ASSCIVRotf_2021,Botti_Conte_ASSCIVRotfproline_2022} to obtain an estimate of the ZPE energy of the several conformers to be compared with DMC values.\cite{conte_glycine20} Results were often in excellent agreement with DMC ones, sometimes showing a difference of less than 5 cm$^{-1}$.
Further details may be found in ref. \citenum{conte_glycine20}.

\section{Gas-phase reaction potentials}

This section examines several MLPs describing chemical reactions in the gas phase with 6 to 9 atoms, so small molecules. 

\begin{figure}[htbp!]
    \centering
    \includegraphics[width=1.0\columnwidth]{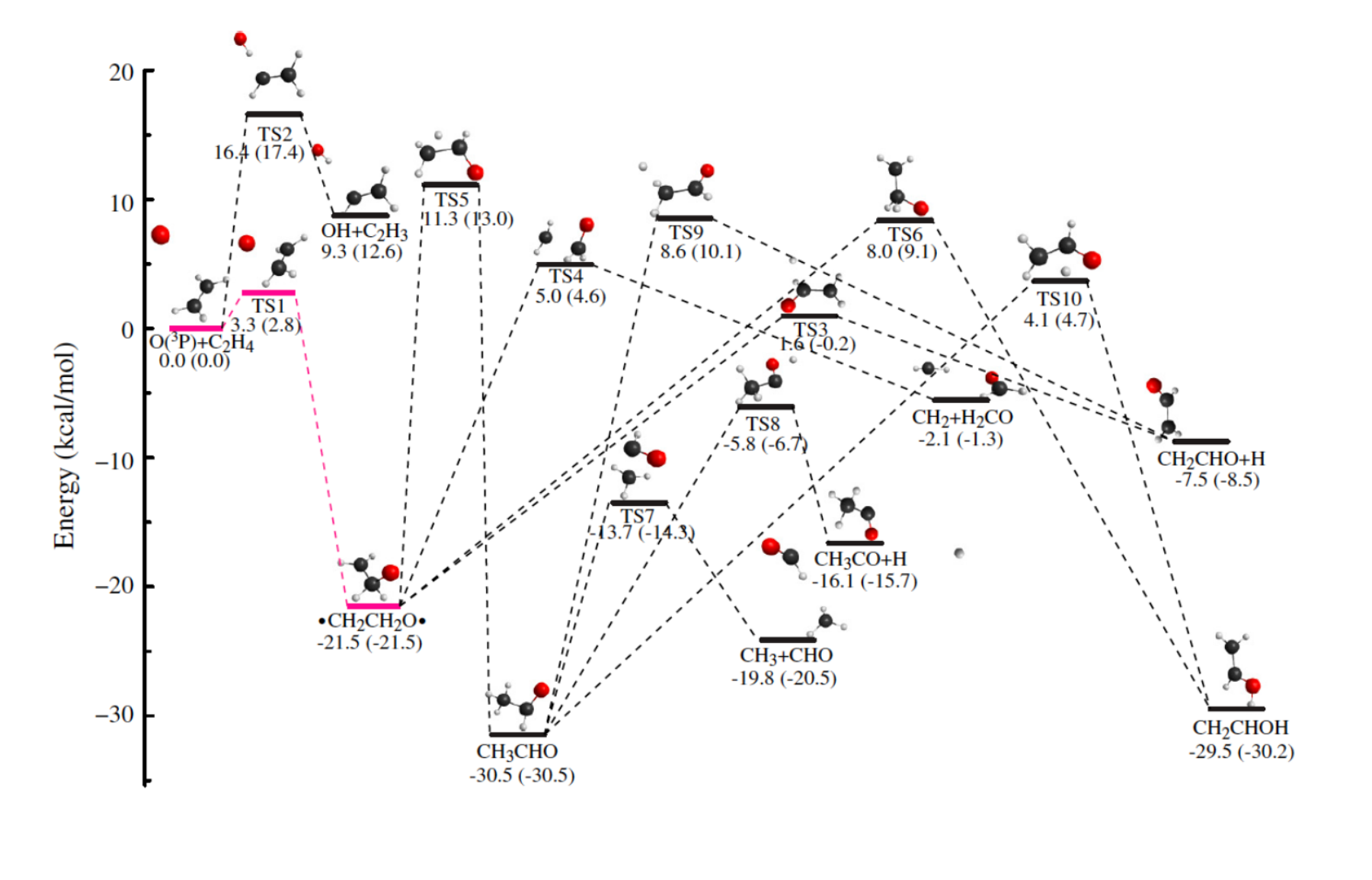}
    \caption{Schematic showing stationary points for the singlet \ce{OC2H4} PIP-PES.  Taken from Ref.\citenum{OC2H4PNAS} The energies given in pairs refer to the PES and (\textit{ab initio}) ones.}
    \label{fig:OC2H4}
\end{figure}

First, consider the singlet PES for acetaldehyde, \ce{CH3CHO}. A schematic of a PIP PES, fit to roughly 200~000 (CCSD(T) plus MRCI) energies \cite{OC2H4PNAS} showing various stationary points and energies is given in Fig. \ref{fig:OC2H4}. (This PES together with one for the triplet state\cite{OC2H4PNAS} were used in QCT calculations in a joint experiment-theory paper.\cite{OC2H4PNAS}) Here we show the singlet PES as it illustrates the complex landscape of the PES. The dataset used in the PES spanned the energy range up to roughly 140 kcal/mol and  all the stationary points. Clearly this is a complex energy landscape that required an extensive dataset in energies and configurations.

\begin{figure}[htbp!]
    \centering
    \includegraphics[width=1.0\columnwidth]{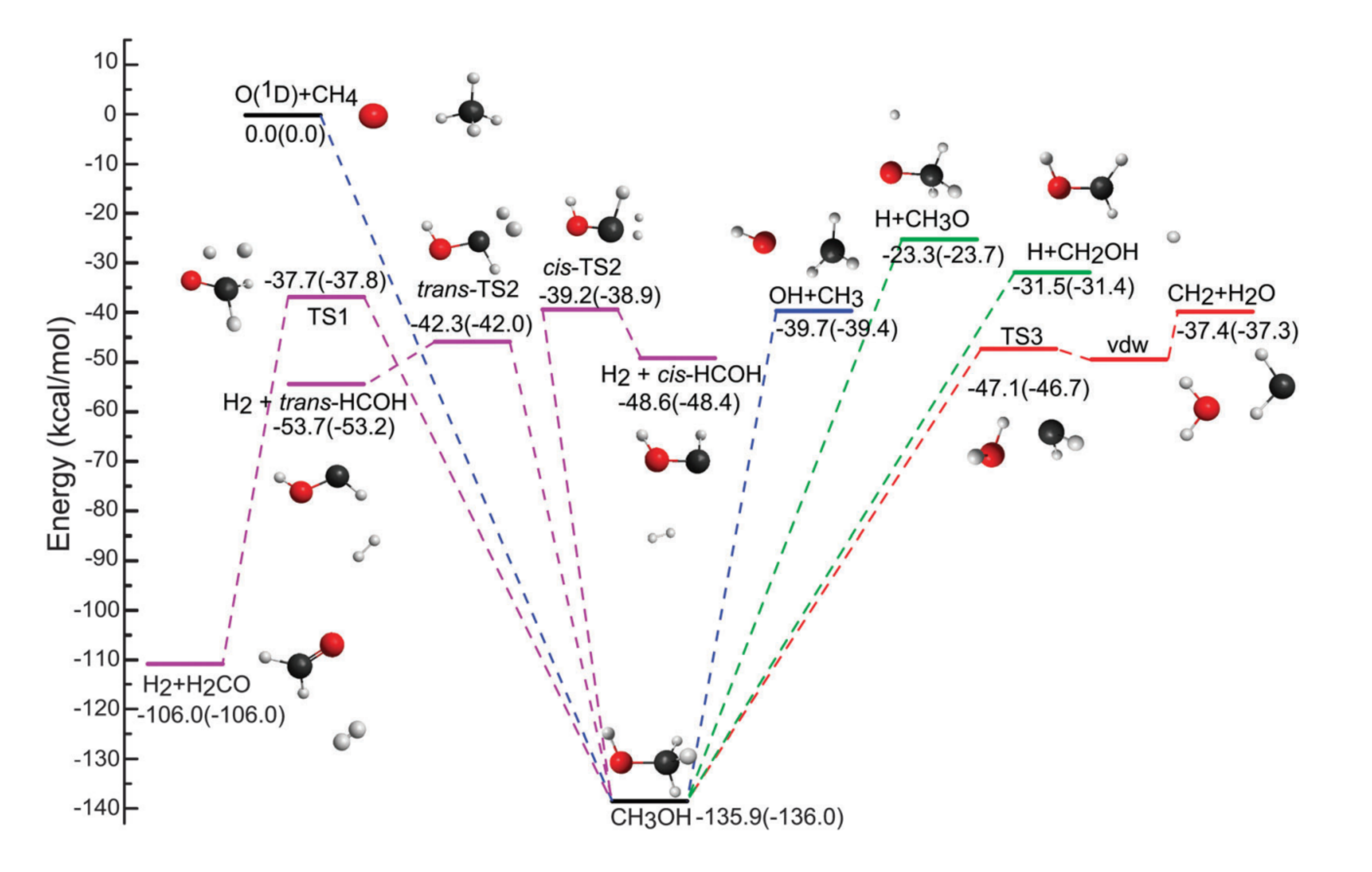}
    \caption{Schematic showing stationary points on the MLP for the reaction O($^1$D)+\ce{CH4}. Taken from Ref. \citenum{OCH4}. The energies given in pairs refer to the PES and (\textit{ab initio}) ones.}
    \label{fig:OCH4}
\end{figure}

An analogous schematic is shown for the recent MLP describing the 6-atom O($^1$D)+\ce{CH4}reaction\cite{OCH4} shown in Fig. \ref{fig:OCH4}. Even though this is a smaller number of atoms, the landscape is complex, and  the dataset, which consisted of roughly 340 000  MRCI + Q/aug-cc-pVTZ electronic energies, is both large and extensive.  This PES was also used in extensive QCT calculations.\cite{OCH4}

\begin{figure}[htbp!]
    \centering
    \includegraphics[width=1.0\columnwidth]{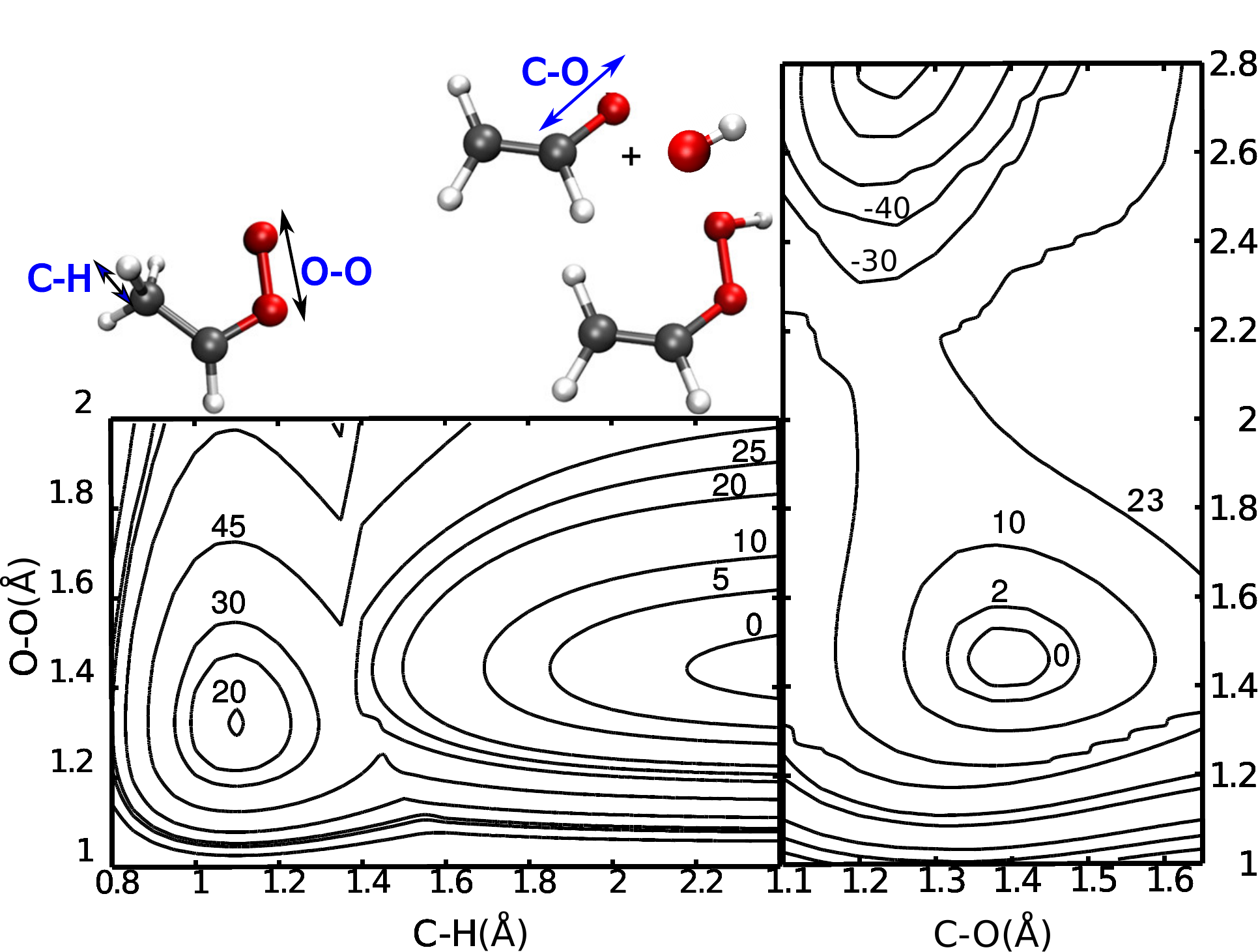}
    \caption{ Two-dimensional cuts through the full-dimensional PES for the $syn$-\ce{CH3CHOO} $\rightarrow$ VHP $\rightarrow$ \ce{CH2CHO}+OH decomposition. The PES is relaxed on each grid point. For the first step (left panel) the CH (proton transfer) and OO coordinates are the driving coordinates whereas for the second step the CO and OO separation are used, respectively.  All energies are given in kcal/mol and the minimum of the VHP state is the zero of energy. The left hand panel represents the hydrogen transfer in $syn$-\ce{CH3CHOO} to form VHP. See text for more details.}
    \label{fig:MM}
\end{figure}

The next example is for the unimolecular dissociation of $syn$-Criegee($syn$-\ce{CH3CHOO}) to OH+\ce{CH2CHO}. This is an 8-atom reaction for which we developed a PIP-PES using a dataset of  157 278 energies (CCSD(T) plus MRCI) up to 70 kcal/ mol relative to the $syn$-\ce{CH3CHOO} minimum.\cite{CH3CHOO}.  Very recently Meuwly and co-workers reported a PhysNet PES for this reaction\cite{CriegeeMM} and used it in trajectory calculations. The fit was trained on a dataset of 84 322 MP2 energies and associated forces.  As with the datasets for the above reaction PESs, this data set contained a large sample of configurations that spanned the reactant, products, stable intermediates including the important vinyl hydroperoxide (VHP), and van der Waals complexes.  Contour plots of important parts of the PES are given in Fig. \ref{fig:MM}. Note beyond the VHP minimum the CH and OO separations are coupled and not useful as the progression coordinate. The right-hand panel shows the O-O bond breaking in VHP. O-O bond-breaking involves multi-reference character which leads to further lowering of the energy beyond the transition state at an O-O separation of 2.2 \AA. This PES is reliable for determining rates but not for final state distributions of the OH product.

\begin{figure}[htbp!]
    \centering
    \includegraphics[width=1.0\columnwidth]{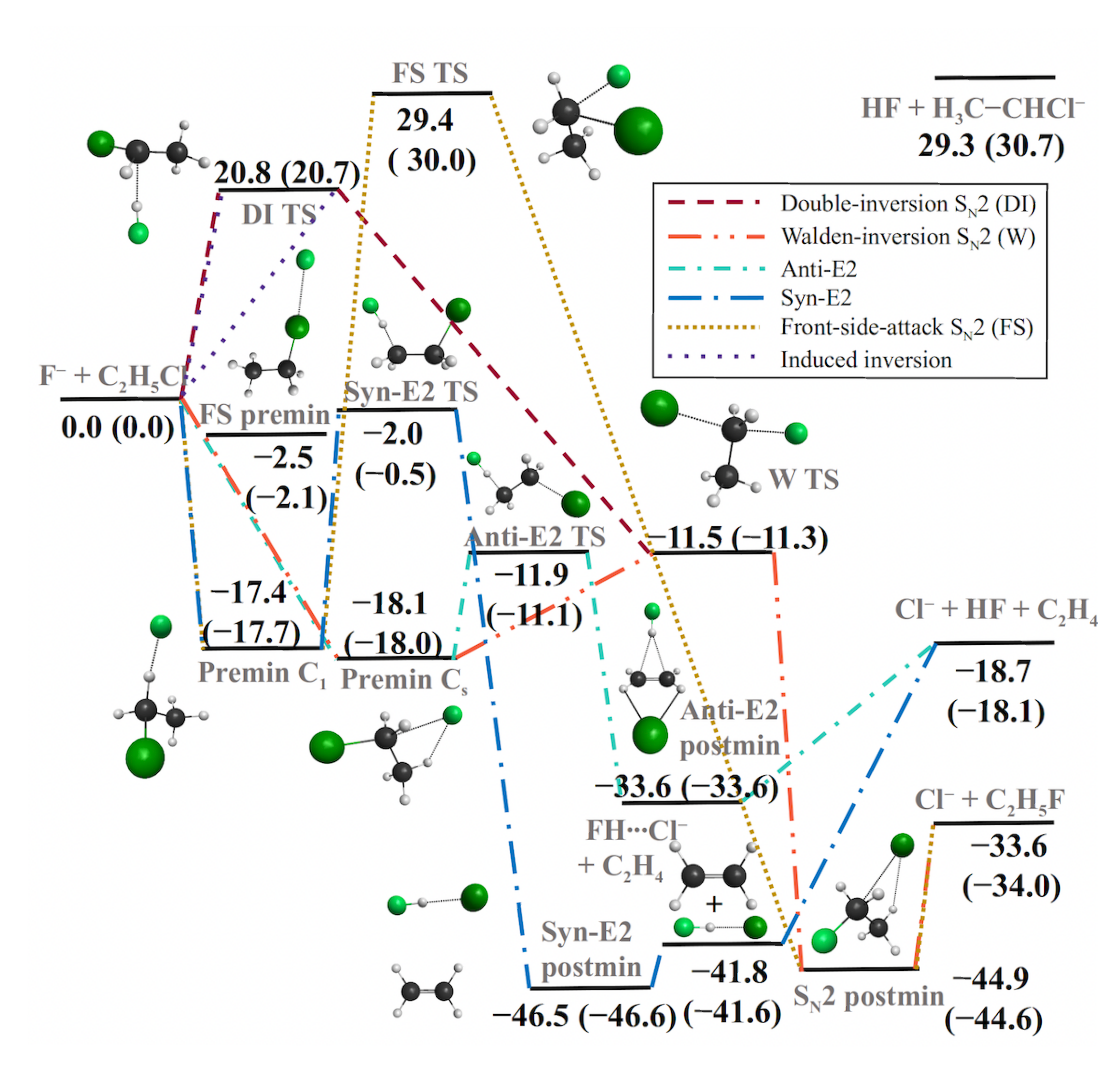}
    \caption{Schematic showing stationary points on the ML PES for reaction F$^-$+\ce{CH3CH2Cl}. Taken from Ref. \citenum{Czako2022}. The energies given in pairs refer to the PES and (\textit{ab initio}) ones.}
    \label{fig:CG}
\end{figure}

The final example is the 9-atom reaction \ce{F^{-} + CH3CH2Cl}.  The details of an MLP for this reaction were recently reported\cite{Czako2022} and used in joint theory (QCT) and experimental study.\cite{CzakoNatChem} The PES was fit to 35 474  CCSD(T)-F12b energies.  The complexity of the PES, i.e., the numerous stationary points and labels of each, is shown in Fig.\ref{fig:CG}. Details of the labels of the stationary points for the interested reader can be found in ref. \onlinecite{Czako2022}.   Clearly the dataset for this MLP is extended in energy and configuration space, as were the datasets shown above.  The relatively  small size of the final dataset was certainly a major accomplishment of this recent work, and more details are given in ref.\citenum{Czako2022}.

\section{Discussion}

To begin this section we return to the MD17 datasets.  As noted these are obtained from direct DFT dynamics at 500 K and so the energy distribution is a single classical one at this temperature. For a molecule of $N$ atoms the average of this distribution is, in the harmonic approximation, ($k_B$T=500/2 K)(3$N$-6)$\approx$ 0.5(3$N$-6) kcal/mol. So for malonaldehyde, ethanol and glycine these averages are 10.5, 10.5 and 12 kcal/mol, respectively. These are all in accord with the classical distributions shown for the MD17 datasets.  

This approach to generate datasets is certainly a reasonable way to create numerous datasets for a series of molecules which may be used primarily for testing and benchmarking MLP methods. Indeed that is primarily what that dataset has been used for. The resulting MLPs can certainly be used in classical MD simulations at thermal energies.  But they would almost certainly not be suitable for rigorous quantum, semi-classical and probably even QCT calculations.  In contrast, the datasets we have generated are molecule-specific and account for specific target properties.    This was described in detail for ethanol, malonaldehyde and glycine. The datasets for ethanol and glycine, like those in MD17, use DFT direct-dynamics for both energies and forces.  However, they are run at several total energies, all much higher than 500 K, and in the example of glycine starting from the various conformers. The different starting conditions are responsible for the shapes seen, i.e., multiple maxima, in the energy distributions shown above. This sampling includes data at or near the various saddle points separating the eight low-lying conformers. These datasets (and the ones for ethanol, malonaldehyde and many other molecules) had quantum DMC calculations of the zero-point state as one target. Indeed, as stated in a recent review ``Nuclear Quantum Effects Enter the Mainstream",\cite{NQE2018} the need for MLPs that are robust for quantum effects is well established. And even for MD calculations of reactions, MLPs require big  datasets.   

The range of nuclear configurations is an important aspect of the PES because this is determined by  the target of the PES.  Targets can range from spanning a single minimum of a PES to describe anaharmonicty and mode-coupling to studies that require a treatment of large amplitude motion perhaps across several minima to the extreme of reaction dynamics with many reaction channels, minima and saddle points. There is a rich history of such studies in the gas phase for small molecules.  In addition, many reactions involve numerous H atoms. Of course, the light H atoms present challenges ranging simply from large zero-point energies to massive tunneling effects, and these are additions to the complexity of the MLPs, especially for reactive MLPs.  Quantum dynamics, which is ideally the approach of choice, makes the most demands on MLPs.\cite{Fu18, guo20}  For reactions with more than 5 atoms the challenges of rigorous quantum calculations are currently insurmountable and so (quasi)classical molecular dynamics calculations are done instead using complex PESs. There are numerous examples of such calculations using robust MLPs and some recent reviews can be found in refs. \citenum{bowman11}, \citenum{CzakoSci11}, and ]\citenum{ EGmode16, cbfeature, czakopers21,Meuwly2022}.

In addition to quantum approaches and their demands on PESs, semi-classical (SC) methods, based on trajectories run on the PES, are able to reproduce quantum effects and also require an accurate description of the PES in the high energy region.\cite{whm2001,Heller1981, Herman_Kluk_SCnonspreading_1984, Kaledin_Miller_Timeaveraging_2003, Aieta_Ceotto_SCnuclDens_2020} This is because SC methods respect the zero-point energy and so dynamics are run at least at this energy.

Semi-quantum methods that benefit from high-energy regions of the PES are the path integral molecular dynamics (PIMD)\cite{Ceriotti_Manolopoulos_PIMD_2010} approach and methods related to it such as centroid molecular dynamics (CMD)\cite{Hone_Voth_CMD_2005, Hele_Althorpe} and ring polymer molecular dynamics (RPMD).\cite{Richardson_Althorpe_Ringpolymer_2009, Habershon_Miller_Ringpolymer_2013} 

To conclude this section, we comment on using direct-dynamics to generate datasets.  This is a good approach as it can explore the configuration space based on the dynamics.  However, one disadvantage is that the size of the data generated in the approach can be both enormous and highly correlated.  So, we generally keep data only for every 10th or so time step for the dataset. This is somewhat wasteful and of course is limited by classical sampling.  So we propose another approach that has the DMC zero-point energy specifically in mind as a target.  The suggestion is inspired by Fig. \ref{fig:histofenergies}, the distribution of potential energies from the DMC wavefunction (represented by thousands of ``walkers'').  Clearly, this would be a good distribution to use for a dataset, if it could be obtained without the PES.  Of course that's not possible exactly.  However, it could be done approximately using an approximate PES. Several possibilities for this range from using a separable harmonic-oscillator model (perhaps ``morsified" to create a separable anharmonic PES) to a model from a force field. We plan to investigate this proposal in the near future.

\section{Summary}
This Perspective has focused on energy datasets for Machine-Learned potentials for ``small" molecules.  We have examined several from the MD17 dataset, which uses an approach that borrows heavily from the perspective of materials research. These have been contrasted with more extensive datasets for the same molecules from the perspective of isolated molecule gas-phase chemical physics. MLPs for reactions of small molecules were also considered.  In both cases we have shown that datasets for small molecules often need to be much larger, both in energies and in configuration space, than those in the MD17 datasets. As such these more extensive datasets, which are available, could provide additional testing datasets for the numerous ML methods. These datasets consitute a new database, QM-22, which currently consists of the datasets listed in Table \ref{tab:Datasets}. The selection includes the datasets for the methane, ethanol, malonaldehyde and glycine PESs discussed here.  The ten additional datasets are recent ones from our group and include some that contain gradient data and those based on CCSD(T) energies which do not.  They all have been used in DMC calculations and thus are ``DMC certified". The protocol to generate each dataset is molecule-specific and is aimed at specific scientific goals. The interested reader is referred to the original paper reporting the dataset for details of how the dataset was generated.
\begin{table}[htbp!]
\caption{Datasets\tnote{a} of indicated molecules in the QM-22 database, available for download.$^a$\tnote{a} }
\label{tab:Datasets}

\begin{threeparttable}
	\begin{tabular*}{\columnwidth}{@{\extracolsep{\fill}}lcc}
	\hline
	\hline\noalign{\smallskip}
     Molecule &     Energies &   Gradients   \\
	\noalign{\smallskip}\hline\noalign{\smallskip}
    Hydronium \ce{H3O+}  &CCSD(T) &no  \\
    \ce{H2CO}, $cis$ and $trans$-HCOH  &   MRCI &no  \\
    Methane \ce{CH4}  &B3LYP &yes  \\
    \ce{OCHCO+} \  &CCSD(T)& no   \\
     Malonaldehyde \ce{C3O2H4} \  &CCSD(T)& no   \\
    Acetaldehyde (singlet) \ce{CH3CHO} \  &CCSD(T)-MRCI& no   \\
    Acetaldehyde (triplet) \ce{CH3CHO} \  &CCSD(T)& no   \\
    $syn$-Criegee \ce{CH3CHOO} \  &CCSD(T)-MRCI& no   \\
    Ethanol \ce{CH3CH2OH}  &B3LYP& yes \\
    Formic acid dimer \ce{(HCOOH)2} &CCSD(T)&  no \\
    Glycine \ce{C2H5NO2}  &B3LYP& yes  \\
    $N$-methyl acetamide \ce{C3H7NO} &B3LYP  & yes  \\
    Tropolone \ce{C7H6O2}  &B3LYP& yes  \\
    Acetylacetone \ce{C5H8O2}  &MP2& yes  \\
	\noalign{\smallskip}\hline
	\hline
	\end{tabular*}
	 \begin{tablenotes}
   \item[a] https://github.com/jmbowma/QM-22
   \end{tablenotes}
\end{threeparttable}
\end{table}

Perhaps this Perspective on ``small" molecule-large datasets is the flip side of ``large" molecules-small datasets, where hopefully the distinction between ``large" and ``small" datasets is clear.

Datasets for molecules that range from 4 to 15 atoms from our group (and one for acetylacetone from the Meuwly group) are publicly available and we hope will be used in new tests of the many ML methods mentioned in the Introduction.

\section*{Acknowledgment}
JMB thanks NASA (80NSSC20K0360) for financial support. RC thanks Universit\`a degli Studi di Milano ("PSR, Azione A Linea 2 - Fondi Giovani Ricercatori") for support. QY thanks Professor Sharon Hammes-Schiffer and National Science Foundation (Grant No. CHE-1954348) for support. We also thank  G{\'a}bor Czak{\'o}, Bina Fu, and Markus Meuwly and Meenu Upadhyay for figures used here. Finally we thank the reviewers, the editor, Markus Meuwly and Matthias Rupp for helpful comments.\

\section*{Data Availability}
Datasets for all molecules listed in Table \ref{tab:Datasets} are  available at
https://github.com/jmbowma/QM-22

\bibliography{refs.bib}
\end{document}